\documentclass[aps,prb,floatfix,twocolumn,showpacs,showkeys]{revtex4-2}
\usepackage{amsmath}
\usepackage{graphicx}
\usepackage{amsfonts}
\usepackage{amssymb}
\usepackage{bm}
\usepackage{epsfig,float,afterpage}
\usepackage[colorlinks,linkcolor=blue,citecolor=blue,urlcolor=blue]{hyperref}
\usepackage[usenames,dvipsnames]{color}
\usepackage{lipsum}
\UseRawInputEncoding
\def\beq{\begin{equation}}
\def\eeq{\end{equation}}
\def\bea{\begin{eqnarray}}
\def\eea{\end{eqnarray}}
\def\nn{\nonumber}
\def\ba{\begin{array}}
\def\ea{\end{array}}

\newlength{\sizeonefig}
\newlength{\sizetwofig}
\begin{document}

\author{Syeda Rafisa Rahaman}
\affiliation{Integrated Science Education \& Research Centre, Visva-Bharati University, Santiniketan-731235, India}
\author{Shreekantha Sil} 
\affiliation{Department of Physics, Visva-Bharati University, Santiniketan-731235, India}
\author{Nilanjan Bondyopadhaya} 
\email{nilanjan.iserc@visva-bharati.ac.in}
\affiliation{Integrated Science Education and Research Centre, Visva-Bharati University, Santiniketan-731235, India}
\title{ Interaction and non-Hermiticity controlled transmission in extended Su-Schrieffer-Heeger models }

\begin{abstract}

We study the transport characteristics of an extended version of the Su-Schrieffer-Heeger (SSH) model with next-nearest-neighbor (NNN) interactions and non-Hermitian onsite energies. We observed that transport in such a system is significantly modified by the NNN interaction and the non-Hermitian terms. The transmission coefficient exhibits oscillatory behavior as the strength of the NNN interaction varies in a fixed-length chain. Moreover, the transmission coefficient also shows oscillation with system size for a fixed strength of the NNN interaction. We find that novel oscillatory behavior of the transmission coefficient, arising form the NNN interaction, is a unique feature of such a model and has not been reported previously. The presence of the non-Hermitian terms also enhances/reduces the transmission coefficient depending on the values of the other system parameters like intra-, inter- and NNN hopping. It appears from our study that both the NNN interaction and the non-Hermiticity introduce significant changes in the transport properties of the extended SSH chain, which are not observed in the standard Hermitian nearest-neighbour variant of the SSH model.

\end{abstract}

\vspace{0.5cm}


\maketitle


\section{Introduction}

The roles of long-range interactions and non-Hermitian extension of Hamiltonian in quantum mechanics have drawn considerable attention in recent times due to their relevance in real experimental scenarios, as well as the existence of exotic phases of matter that originate from these two features. Long-range interactions, which occur in several quantum many-body systems, significantly influence the energy spectrum and the topological properties of the systems \cite{Degottardi_PRB,Viyuela_PRL,Defenu_RMP_2023}, and these, in turn, affect the transport characteristics of the systems \cite{Gonzalez_PRB, Chavez_PRL, Banerjee_2024}. In recent times, realization of long-range interactions has been achieved in several experimentally accessible platforms, namely Rydberg atom arrays \cite{Choi_Science,Saffman_RMP}, magnetic impurities induced sub-gap Shiba states inside s-wave superconductors \cite{PientkaPRB2013,PientkaPRB2014}, an array of atomic ensembles within an optical cavity \cite{Pariwal2021}, polar molecule simulators \cite{Carr_NJP_2009,Lahaye_RPP_2009} and trapped ions \cite{Haffner_PR}. However, in our analysis, we concentrate on the NNN type of interaction which is the simplest type of long-range interaction and could be achieved in tunable setups with a controllable interaction range.

Non-Hermitian quantum systems yield a wide range of fascinating results distinct from those of conventional Hermitian quantum systems \cite{Bender_2007,Elganainy_NatP,Ashida_AdvPhy}. A class of non-Hermitian systems is characterized by the presence of gain and loss terms in the Hamiltonian in which gain and loss terms give rise to the non-Hermiticity. Inclusion of non-Hermitian terms in the Hamiltonian can be regarded as an efficient way to describe the open quantum systems in which dissipation or fluctuation-induced dissipation is the physical origin of non-Hermiticity \cite{Rotter_2009,Wang_CPB}. Dissipation which can arise from different physical mechanisms, namely, proximity effect in heterostructures \cite{Chuan_PRB}, cold atoms \cite{Li_Natcom}, photonic lattices \cite{Ozde_NatM}, superconducting circuits \cite{Blais_RMP, Fitz_PRX}, has significant effects on quantum transport \cite{Yang_PRL}.

The Su-Schrieffer-Heeger (SSH) chain is a simple model of a topological insulator in one dimension that hosts edge states in a non-trivial topological phase provided that "intra-cell" hopping is less than "inter-cell" hopping \cite{SuPRL,HeegerRMP}. The SSH model is a well-studied model from both theoretical \cite{Asboth2016,wang_PRB} and experimental  perspectives. Interestingly, it has been realized experimentally in several systems, namely, mechanical systems \cite{chaunsali_PRL}, an array of inductor-capacitor (LC) waveguides \cite{KimPRX}, ultracold bosonic atoms in 1D superlattice \cite{Grusdt_PRL}, Rydberg synthetic lattice of ${}^{84}Sr$ \cite{Kanungo_NatCom} and ${}^{87}Rb$ \cite{Meier_Natcom,Meier_Science}, Hardcore bosons \cite{Sylvain_Science}, 1D lattice of polariton micropillars \cite{Stjean_NatPho}, nanoparticles \cite{kruk_small}, supercondutors \cite{Cai_PRL}, silicon quantum dots \cite{Simmons_Nat}, artificial lattice made of Cs/InAs(111)A \cite{Ligthart_PRR}, 1D trapped-ion chains with tunable interaction range \cite{Katz_NatCom} etc. Among these, some setups are controllable in which the hopping strength and onsite energy can be tuned to drive the system between topological and non-topological phases. From an experimental point of view, obviously, realizing a generalized SSH chain in a laboratory setup requires a higher level of engineering as it involves controlling the next-nearest-neighbor interactions and onsite energy. The extended SSH chain with long-range interaction has been realized in some laboratory experiments, such as ultracold atoms in a momentum lattice \cite{Dizhou_njp}, 
optically resonant nanoparticles \cite{Li_physics}, photonic crystal systems \cite{Li_PRB2014}, and electric circuit \cite{Yin_2025}. 

In this article, we consider an extended SSH chain with next-nearest-neighbor (NNN) interactions \cite{Li_PRB,SilPRB2025} and non-Hermitian complex onsite energy terms \cite{SilPRB2025,LonghiOL2018}. Evidently, it is a generalized variant of the SSH chain as it includes NNN interaction and non-Hermitian onsite energy terms. The presence of two such special features, which are beyond the scope of the ordinary SSH model, provides a platform for investigating the interplay between these two marked characteristics. 
We investigate the transport properties of the extended SSH chain while varying the NNN interaction strength, length of the chain and non-Hermitian onsite potentials. We observe strong modulation of the transmission properties of the extended SSH chain in the following two cases: (i) as the strength of the NNN interaction varies for a fixed-length chain, and (ii) as the length of the chain varies for a fixed non-zero NNN interaction. In addition, we notice that the non-Hermitian onsite energy affects the transmission properties of an extended SSH chain, which is also observed in the case of the Hatano-Nelson-type tight-binding model in Ref. \cite{TakaneJPSJ_2023}. 

The article is organized as follows: In Sec. \ref{sec:Hamiltonian}, we discuss the Hamiltonian of the extended SSH chain with NNN interactions and non-Hermitian onsite energy terms. We present and explain all all the results obtained from the numerical simulation in Sec. \ref{sec: Results}. This section is divided into three subsections in which we study the modulation of the transmission coefficient under the influence of three tunable parameters of the extended SSH chain, namely the strength of the NNN interaction, the system size and the non-Hermitian onsite energy terms. Finally, we conclude in Sec. \ref{sec:con} after a brief summary. In the Appendix. \ref{App1}, we discuss the scattering matrix approach that has been employed for the calculation of the transmission coefficient in this article. 

\section{Model Hamiltonian and Formalism}
\label{sec:Hamiltonian}

\begin{figure}[htb]
\centering{\includegraphics[width=0.8\linewidth]{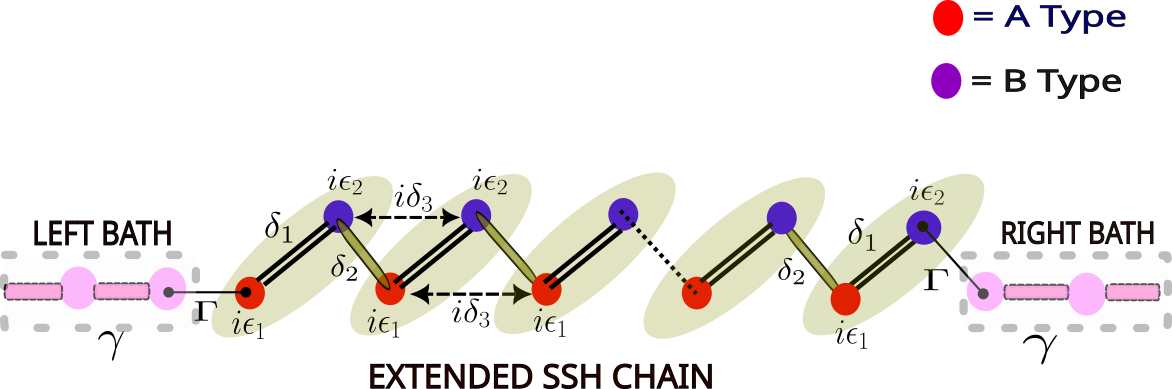}}
\caption{Extended SSH chain connected with baths.}
\label{cartoon}
\end{figure}

We consider a one-dimensional extended SSH chain with unit cells made of two types of $A$ and $B$ sites. The Hamiltonian of an extended SSH chain with $N$ unit cells on a lattice of $2N$ sites is given by 
\bea
H_{S}&=& \left(\delta_1 \sum_{m=0}^{N-1}c_{2m+1}^\dagger c_{2m+2} +\delta_2 \sum_{m=1}^{N-1}c_{2m}^\dagger c_{2m+1} + \text{H.c.} \right) \nn \\
&&+\left( i \delta_3 \sum_{m=1}^{2N-2} c^\dagger_{m+2} c_m +\text{H.c.}\right) \nn \\
&&+ i \left( \epsilon_1 \sum_{m=0}^{N-1} c_{2m+1}^\dagger c_{2m+1}+\epsilon_2 \sum_{m=1}^{N}c_{2m}^\dagger c_{2m}\right) \, ,
\label{ESSH}
\eea
where, $c_{2m+1}^\dagger, m \in \{0,1,\dots,N-1\} $ ($c_{2m}^\dagger, m \in \{1,\dots,N\}$) are creation operators localized at $A$ ($B$) type sites. Here, $\delta_1$ and $\delta_2$ are hopping amplitudes from $A \rightarrow B$ (intra) and $B \rightarrow A$ (inter), respectively; $\delta_3$ is the next-to-nearest neighbor hopping strength between two adjacent $A$/$B$ type sites from two adjacent unit cells. 
This generalized model also has non-Hermitian terms with strengths $\epsilon_1$ and $\epsilon_2$ for $A$ and $B$ type sites, respectively (See Fig. \ref{cartoon}). In case both $\epsilon_1$ and  $\epsilon_2$ have the same sign, these non-Hermitian terms correspond to purely loss or gain terms, where as for $\epsilon_1=-\epsilon_2$, these terms represent balanced gain-loss terms.

In order to study transport in this open chain, we connect the same with two identical semi-infinite Hermitian leads. The left lead extended to $m \rightarrow -\infty$ is represented by $\mathcal{H}_L$, and the right lead extended to $m \rightarrow \infty$ is denoted by $\mathcal{H}_R$. $\mathcal{H}_L$ and $\mathcal{H}_R$ are connected to the left and right ends of the middle SSH chain, respectively.

\bea
\mathcal{H}_L &=&\gamma \sum_{m \leq 0} \left( c^{\dagger}_{m+1}c_m +c_m^\dagger c_{m+1} \right) \, , \nn \\
\mathcal{H}_R &=&\gamma \sum_{m \geq 2N } \left( c^{\dagger}_{m+1}c_m +c_m^\dagger c_{m+1} \right) \, ,
\eea
where $\gamma$ is the hopping strength of both leads. These leads are connected to the middle wire via couplings Hamiltonian, 
\bea
\mathfrak{H}_L&=&\Gamma (c^{\dagger}_0 c_1 + c^\dagger_1 c_0) \, , \nn \\
\mathfrak{H}_R&=&\Gamma (c^{\dagger}_{2N} c_{2N+1} + c^\dagger_{2N+1} c_{2N}),
\eea
respectively for left and right junctions, and $\Gamma$ is the tunnel coupling, which is the same for both junctions. Therefore, the full Hamiltonian involving the extended SSH chain, leads and couplings reads
\beq
H=H_S+\mathcal{H}_L +\mathcal{H}_R + \mathfrak{H}_L+\mathfrak{H}_R\, .
\eeq

We use scattering matrix theory to determine the transmission ($T$) and Reflection ($R$) coefficients for the transport. Following Ref. \cite{TakaneJPSJ_2023,Ando_PRB}, we assume that an incoming plane wave $\psi^{IN}(m)=e^{i k m a}\,(m \leq 0\, \rm{and} \, a= \rm{lattice \,constant\, of \, lead})$ is incident on the left edge ($m=1$ site) of the generalized SSH chain from the left lead. Here, the momentum of the plane wave is specified by a real number $k$, and its energy is specified by the dispersion relation of the left lead, which reads $E=2 \gamma \cos(ka)$. Upon incident, the incoming plane wave,  $\psi^{IN}(m)$, is divided into two parts, one part with amplitude $\tilde{r}$ is reflected back to the left lead, and another part enters the middle wire. Therefore, the position space wave functions for the $m$-th site of the left lead are given by 
\beq
\psi_L(m)=e^{i k a m}+\tilde{r}e^{-i k a m}\, ,
\eeq
where $m \leq 0$. Plane waves which enter the middle wire are scattered by the extended SSH chain. We denote the position space wave function for the $m$-th site of the middle site as $\psi_C(m)$, where $ 1 \leq m \leq 2N$. This part of the wave function contains both left-moving and right-moving, which superpose and cause interference. Clearly, this interference is largely influenced by the strengths of $\delta_1$, $\delta_2$ and $\delta_3$, making the scattering process nontrivial. Since the scattering process is elastic inside the middle wire, the energy of the incident wave remains the same everywhere in the system. Finally, the fraction of the plane wave that comes out from the last site, i.e. the $2N$-th site of the middle wire, is transmitted to the right lead. If $\tilde{t}$ is the transmission amplitude of the outgoing wave, the right bath wave function can be written as
\beq
\psi_R(m)=\tilde{t} \, e^{i k a [m - (2N+1)]} \, ,
\eeq
where $m \geq 2N+1$. Hence, the full state vector representing the whole system reads
\beq
| \psi \rangle = \sum_{m=-\infty}^0 \psi_L(m)|m \rangle + \sum_{m=1}^{2N} \psi_C(m) |m \rangle +\sum_{m=2N+1}^{\infty} \psi_R(m)|m \rangle \, ,
\eeq
where $\{ |m \rangle  \}$ are the basis vectors of the position space of the full lattice consisting of leads' sites and SSH chain's sites. 

In general, system state vectors satisfy the Schr\"{o}dinger equation: $H |\phi(E) \rangle = E | \phi(E) \rangle$, where $E$ is the energy eigenvalue of the energy eigenstate $|\phi(E)  \rangle$. For a plane wave state, Schr\"{o}dinger equation reads,
\beq
H |\psi \rangle = E |\psi \rangle \, ,
\eeq
where, $E= 2 \gamma \cos(ka)$ is the energy of the incoming plane wave with momentum $k$. 
We use scattering matrix theory to calculate the transmission coefficient for the outgoing waves from the extended SSH chain. According to our construction, it is the transmission coefficient ($T$) at the $(2N+1)$-th site of the full lattice, hence it is defined by 
\beq
T = |\psi_R(2N+1)|^2= 4 \gamma^2\,| [G]_{2N+2,1} \sin (ka)|^2\, ,
\eeq
where $ G$ is the matrix Green's function (See Appendix~\ref{App1} for the detailed form). 


\section{Results and Discussions}
\label{sec: Results}
The results section is divided into three subsections, in which the first two subsections (\ref{subsec: A}, \ref{subsec: B}) deal with the Hermitian variant of the extended SSH ($\epsilon_1=\epsilon_2=0$) and in the third subsection (\ref{subsec: C}), we consider the full non-Hermitian version of the same ($\epsilon_1\neq 0, \epsilon_2 \neq 0$). First, we study the variation of transmission coefficient ($T$) with $\delta_3$ for a fixed-length chain. Secondly, we study the variation of $T$ as a function of chain's length for a fixed value of $\delta_3$. Finally, we investigate the effect of non-Hermiticity on the $T$. This is worth noting that in the following sections, we represent all other coupling parameters in the scale of $\gamma$ while choosing $\Gamma=\gamma=1$. Further, we set $a=1$ hereafter without loss of generality.

\subsection{Variation with NNN interaction strength}
\label{subsec: A}
\begin{figure}[htb]
\centering{\includegraphics[width=1\linewidth]{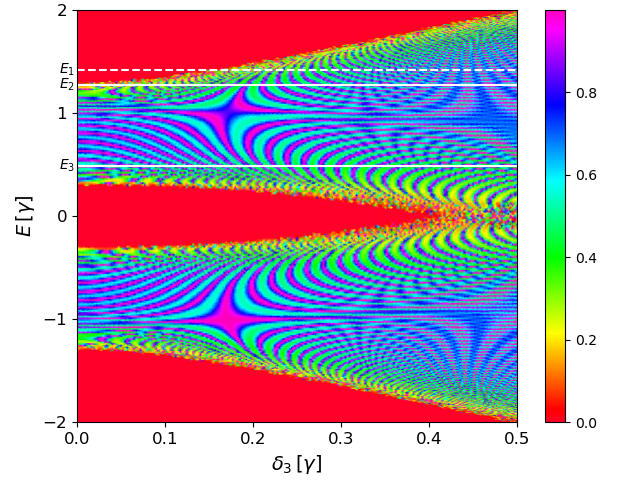}}
\caption{Density plot of $T$ in $E-\delta_3$ plane for a Hermitian variant of extended SSH chain with $N=100$ unit cells; $\delta_1=0.5 \gamma, \delta_2=0.8 \gamma, \epsilon_1=\epsilon_2=0 $.}
\label{density_N100_T_d3}
\end{figure}

\begin{figure}[htb]
\centering{\includegraphics[width=1\linewidth]{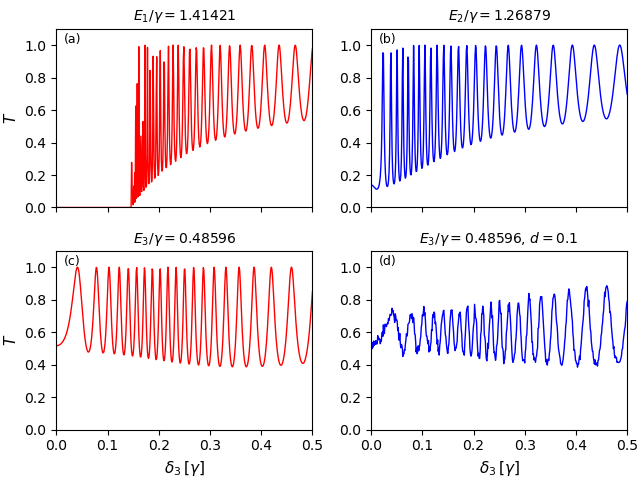}}
\caption{$T$ vs. $\delta_3$ for three different energies $E_1$, $E_2$ and $E_3$; length of the extended SSH chain is $N=100$ unit cells; $\delta_1=0.5 \gamma, \delta_2=0.8 \gamma$. In all subplots, $\epsilon_1=\epsilon_2=0$. }
\label{T_vs_d3_diso}
\end{figure}

We investigate the variation of the $T$ with the strength of the next-nearest-neighbor (NNN) interaction ($\delta_3$), which in this model is considered as a long-range interaction, and the energy of the incoming plane wave ($E$). In Fig. \ref{density_N100_T_d3}, we depict the variation of $T$ with both $\delta_3$ and $E$ for an extended SSH chain with $N=100$ unit cells. Magnitude of $T$ is represented by the by color gradation. $T$ shows interesting variation with $\delta_3$ for different energies ($E$) of the incoming waves. In some specific ranges of $E$, $T$ varies significantly from low to high values as $\delta_3$ changes. This figure shows significant variations of $T$ with $\delta_3$, which is unique to the extended SSH chain in comparison to the standard SSH chain. In order to understand these variations, we choose three different incoming plane waves with energies $E_1$, $E_2$ and $E_3$, respectively. For both  Fig. \ref{density_N100_T_d3} and Fig. \ref{T_vs_d3_diso}, the SSH chain's parameter values are taken as $\delta_1/\gamma=0.5, \delta_2/\gamma=0.8, \epsilon_1=\epsilon_2=0$. 

Three horizontal white lines corresponding to energy values $E_1$, $E_2$ and $E_3$ are shown in Fig. \ref{density_N100_T_d3}. Evidently, the magnitude of the $T$ oscillates as $\delta_3$ changes from $0$ to $0.5$ along the white lines from left to right end. This oscillation of $T$ with $\delta_3$ is plotted in Fig. \ref{T_vs_d3_diso}. Fig. \ref{T_vs_d3_diso} (a), (b) and (c), represent oscillations of $T$ for three different energies $E_1/\gamma=1.41421$, $ E_2/\gamma=1.26879$ and $E_3/\gamma=0.48596$, respectively. In Fig. \ref{T_vs_d3_diso} (a), $T$ becomes non-zero only after $\delta_3/\gamma > 0.145$. As $\delta_3$ increases further, $T$ undergoes oscillation, initially with a large amplitude that gradually decreases in Fig. \ref{T_vs_d3_diso} (a-b), as the strength of long-range interaction increases. We will see later that the near-zero values of $T$ for the initial values of $\delta_3/\gamma \in \{0,0.145\}$ are due to the absence of density of states of the Hamiltonian for weak NNN interaction.
Moreover, as the strength of the long-range interaction increases, $T$ tends to saturate at a maximum value as a result of strong correlations between neighboring lattice sites due to long-range interaction. However, we observe that the $T$ oscillation begins at the very start of the $\delta_3$ variation in \ref{T_vs_d3_diso}  (b-c). This happens because the DOS study of the extended SSH chain reveals the existence of sufficient numbers of low-energy states ($E \in \{ E_2, E_3 \} $)  even for weak NNN interaction. Interestingly, in this subplot (c), the amplitudes of the oscillation show very slight decrements as $\delta_3$ increases. 

We also study the effect of moderate disorders on the oscillation of $T$ in Fig.  \ref{T_vs_d3_diso}  (d). All the parameters except the strength of the disorder are the same in both  Fig.  \ref{T_vs_d3_diso}  (c) and (d), but in Fig. \ref{T_vs_d3_diso}  (d), there are additional onsite random disorders which are randomly distributed in the range $\{-d/2,d/2 \}$ with strength $d=0.1$. This result is an average of $100$ different sets of data with random onsite disorders. It is evident from this figure that the moderate disorders mainly suppress the amplitudes of the $T$ oscillations, while the overall qualitative feature of $T$ oscillation, i.e. the oscillatory behavior of $T$ with $\delta_3$, remains unaltered. Further, in the disorder case, the amplitude of the oscillations increases with $\delta_3$, indicating that the NNN interaction protects the states against disorder.

%

\begin{figure}[htb]
\centering{\includegraphics[width=1\linewidth]{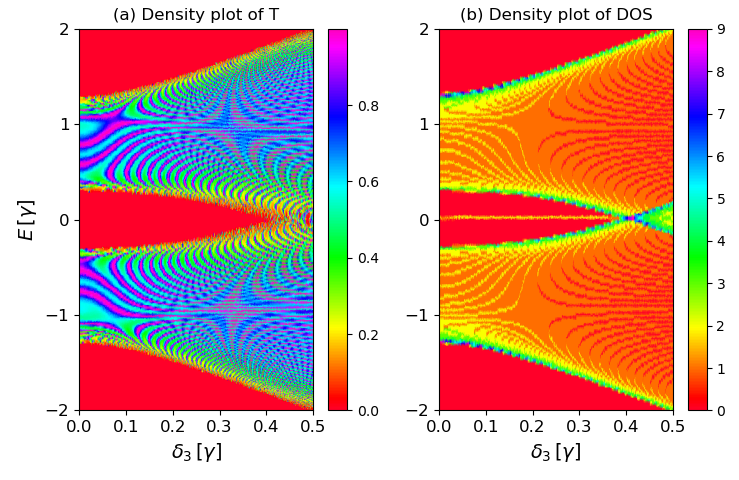}}
\caption{Density plot of $T$ and DOS in $E-\delta_3$ plane for a Hermitian variant of extended SSH chain with $N=75$ unit cells; $\delta_1=0.5 \gamma, \delta_2=0.8 \gamma, \epsilon_1=\epsilon_2=0 $.}
\label{T_DOS_N75}
\end{figure}

To understand the relation between the variation of $T$ and density of state (DOS), we plot both $T$ and DOS in $E-\delta_3$ plane for an extended SSH chain with $N=75$ unit cells. These density plots is depicted in Fig. \ref{T_DOS_N75}. It can be noted that the density plots of $T$ and DOS are not exactly the same, because $T$ is calculated while connecting the SSH chain with leads at both sides; on the other hand, the DOS is calculated for an isolated extended SSH chain Hamiltonian with $N=75$ unit cells. Fig. \ref{T_DOS_N75} (a) represents $T$ variation with both $E$ and $\delta_3$, which also shows oscillatory behavior of $T$ with $\delta_3$ along a straight lines parallel to $x$-axis which corresponds to a incident plane wave with fixed energy $E$. A density plot of DOS of an isolated extended SSH chain is depicted in Fig. \ref{T_DOS_N75} (b), which also shows density oscillations as $\delta_3$ varies for fixed $E$.
 
We find that the transmission coefficient strongly depends on the strength of NNN interaction. As the strength increases, the transmission coefficient undergoes oscillating behavior between $0$ and $1$ values. In all cases, $T+R=1$ as the Hamiltonian is Hermitian. In some regions of parameters space, $T$ is close to $1$, which indicates that the extended SSH chain becomes transparent for these values of parameters. In contrast, for other regions, $T$ becomes minuscule, which implies the opaqueness of the extended SSH chain for incoming plane waves. We think this oscillating behavior of $T$ is a result of the interference between the incident plane waves and reflected plane waves, which are being reflected from the junctions between two adjacent unit cells and the junctions between the extended SSH chain and the leads. However, NNN interaction strength significantly controls the interference pattern. This phenomenon can be interpreted as the interaction-induced transparency in the extended SSH chain. As the strength increases, this NNN interaction effectively reduces the resistive properties of transport and makes the system almost transparent for most of the incoming plane waves with different energies.


%

\subsection{Length-dependent oscillation}
\label{subsec: B}

\begin{figure}[htb]
\centering{\includegraphics[width=1\linewidth]{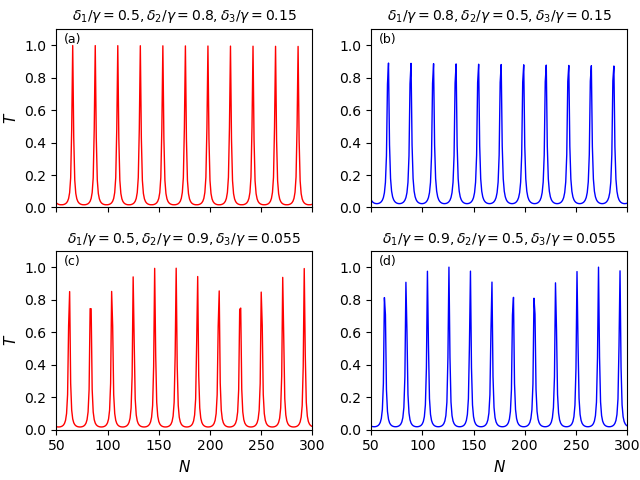}}
\caption{Variation of $T$ with $N$ while $\delta_3$ is fixed and the energy of the incident plane wave is fixed at $E/\gamma=1.41421$. In (a) $\delta_1=0.5 \gamma,\delta_2=0.8 \gamma, \delta_3=0.15 \gamma$, (b) $\delta_1=0.8 \gamma,\delta_2=0.5 \gamma, \delta_3=0.15 \gamma$, (c) $\delta_1=0.5 \gamma,\delta_2=0.9 \gamma, \delta_3=0.055 \gamma$ and (d) $\delta_1=0.8 \gamma,\delta_2=0.5 \gamma, \delta_3=0.055 \gamma$. In all subplots, $\epsilon_1=\epsilon_2=0$.}
\label{T_vs_N_reg}

\end{figure}

The transmission coefficient shows strong length dependence in the presence of the NNN interaction for a fixed energy plane wave. Fig. \ref{T_vs_N_reg} depicts the regular oscillatory behavior of $T$ as a function of the number of unit cells ($N$) or, equivalently, the length of the chain for a fixed energy incident wave. In Fig. \ref{T_vs_N_reg}, $T$ oscillations are very prominent and regular in all the sub-figures. Fig. \ref{T_vs_N_reg} (a) shows the sharp oscillatory behavior of $T$ from $0$ to $1$ in the topological phase $(\delta_1/\gamma=0.5,\delta_2/\gamma=0.8)$ of extended SSH chain with long-range interaction, $\delta_3/\gamma=0.15$. However, the maximum value of $T$ slightly reduces to $0.9$ in the non-topological phase $(\delta_1/\gamma=0.8,\delta_2/\gamma=0.5)$ of extended SSH with the same strength of long-range interaction, i.e. $\delta_3/\gamma=0.15$. For these two sets of $(\delta_1, \delta_2)$, $\delta_3/\gamma =  0.15$ produces $T$ oscillation with maximum amplitudes and almost a single frequency. Similarly, \ref{T_vs_N_reg} (c) and (d), depict simple periodic oscillations of $T$ with the $N$ for $(\delta_1/\gamma,\delta_2/\gamma)=(0.5,0.8)$ and $(0.8,0.5)$, while $\delta_3/\gamma=0.055$. We find that in \ref{T_vs_N_reg} (a)-(d), $T$ values vary from $0$ to $1$ as $N$ changes, for some specific values of $\delta_3$. This NNN interaction-induced oscillation of $T$ between $0$ and $1$ indicates alternate cycles of opaque phases and transparent phases of the extended SSH chain as the length of the chain varies. This is a unique feature of the extended SSH chain, which goes through cycles of opaque and transparent phases as the chain length changes, while the strength of the NNN interaction is tuned at some specific values. 

\begin{figure}[htb]
\centering{\includegraphics[width=1\linewidth]{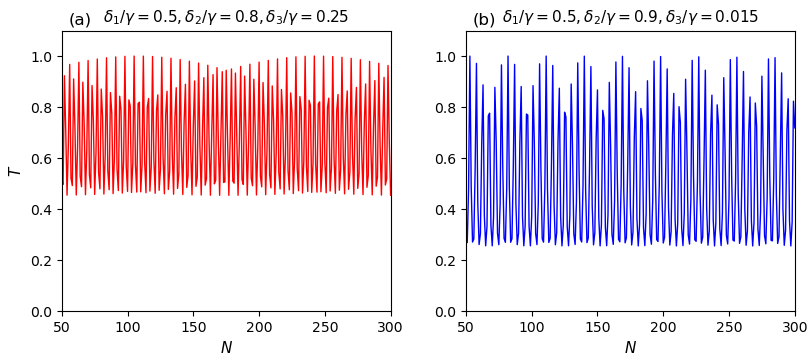}}
\caption{Irregular oscillation of $T$ with $N$ for (a) $\delta_1=0.5 \gamma, \delta_2=0.8 \gamma, \delta_3=0.25 \gamma$ and (b) $\delta_1=0.5 \gamma, \delta_2=0.9 \gamma, \delta_3=0.015 \gamma$. In all subplots, $\epsilon_1=\epsilon_2=0$.}
\label{T_vs_N_irr}
\end{figure}

We also observe that for a given set of $(\delta_1,\delta_2)$, any arbitrary value of $\delta_3$ does not produce such a simple oscillation with large amplitude, rather, most of the values of $\delta_3$ produce a complex oscillatory pattern in $T$ characterized by smaller amplitudes and the superposition of multiple frequencies. Fig. \ref{T_vs_N_irr} depicts such complex oscillation patterns of the transmission coefficient for other values of $\delta_3$, which are chosen arbitrarily. We do not find any significant difference in $T$ oscillations between the topological and non-topological phases of the SSH chain. It is worth noting that in the absence of the NNN interaction, this oscillation of $T$ does not exist at all. 

We verify that the transmission coefficient equals $1$ for all incident plane waves, when $\Gamma = \delta_1=\delta_2$. It is expected because in this limit, the SSH chain connected with the leads behaves like a homogeneous wire, hence no scattering takes place at the junctions. However, while lead strength is not equal to the chain hopping strengths, we find oscillating behavior of $T$ as we change the length of the wire. In such cases, it is possible to define the frequency of such oscillations, which depends on the system parameters. 
%

\subsection{Effect of non-Hermiticity}
\label{subsec: C}
Inclusion of non-Hermitian onsite energy terms significantly modify the transmission property of the extended SSH chain. We consider two different cases of non-Hermitian onsite energy, namely (i) balanced gain-loss terms i.e. $\epsilon_1=-\epsilon_2=\epsilon$ , and (ii) pure loss/gain terms when  $\epsilon_1=\epsilon_2=\epsilon$.

\subsubsection{Balanced gain-loss terms { \normalfont ($\epsilon_1=-\epsilon_2=\epsilon$)}}

The balanced gain-loss terms introduce non-Hermiticity in the extended SSH chain. However, depending on the strength of the balanced gain-loss terms, the extended SSH chain can be in the $\mathcal{PT}$-symmetric phase or the $\mathcal{PT}$-broken phase. For an isolated extended SSH chain with balanced gain-loss terms under the periodic boundary condition (PBC), the dispersion relation of the bulk spectrum can be calculated by taking the Fourier transformation of the Hamiltonian (\ref{ESSH}), and the dispersion relation reads \cite{SilPRB2025},
\beq
E_\pm (k)=-2 \delta_3 \sin k \pm \sqrt{\delta_1^2+\delta_2^2 +2 \delta_1 \delta_2 \cos k -\epsilon^2} \, ,
\eeq
where $k$ is the quantized wave vector under PBC, and $k=\frac{ \pi s }{N} $ for $s=0,1,2, \dots, (2N-1)$. It is evident from the dispersion relation that the bulk energy eigenvalues become real if $|\delta_1-\delta_2| > \epsilon$. On the other hand, energy eigenvalues become imaginary when $|\delta_1-\delta_2| < \epsilon$. However, numerical diagonalization of the extended SSH Hamiltonian with balanced gain-loss in the topological phase ($ \delta_1 < \delta_2 $) reveals that the energies of the edge states always remain imaginary for any non-zero value of $\epsilon$. This Hamiltonian (\ref{ESSH}) shows $\mathcal{PT}$-symmetry provided $|\delta_1-\delta_2| > \epsilon$ condition is satisfied \cite{SilPRB2025}. Nevertheless, when this isolated Hamiltonian is connected with leads via coupling, the characteristics of the energy eigenvalues of the isolated Hamiltonian changes to some extent. Numerical diagonalization shows that the coupling with leads can introduce a few states with imaginary energy even in the $\mathcal{PT}$-symmetric Hamiltonian. 
\begin{figure}[htb]
\centering{\includegraphics[width=1\linewidth]{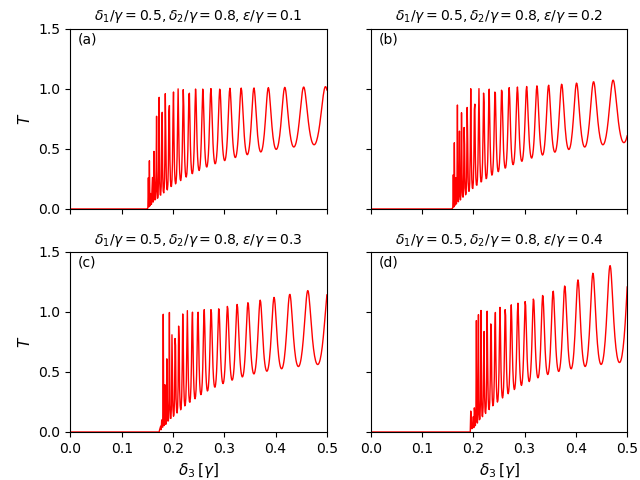}}
\caption{Variation of $T$ with $\delta_3$ for an incident plane wave with $E/\gamma=1.41421$ while the extended SSH chain is in the topological phase ($\delta_1 < \delta_2$). In (a) $\epsilon=0.1\gamma$, (b) $\epsilon=0.2 \gamma$, (c) $\epsilon=0.3 \gamma$ and (d) $\epsilon=0.4 \gamma$. For all subplots $\delta_1=0.5 \gamma$, $\delta_2=0.8\gamma$, and the length of the chain is $N=80$ unit cells.}
\label{T_vs_D_Top}

\end{figure}

\begin{figure}[htb]
\centering{\includegraphics[width=1\linewidth]{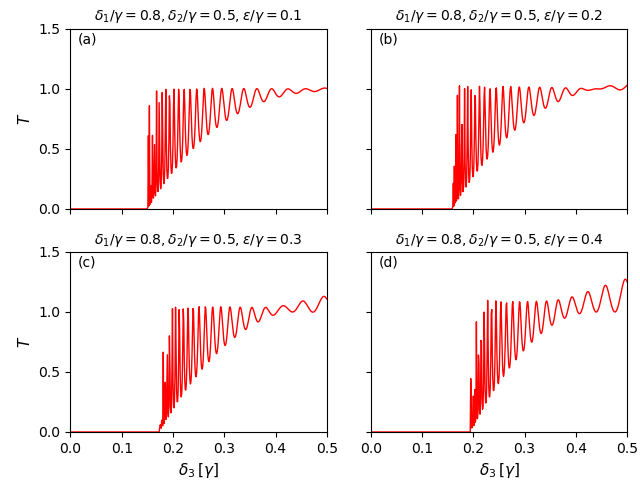}}
\caption{Variation of $T$ with $\delta_3$ for an incident plane wave with $E/\gamma=1.41421$ while the extended SSH chain is in the non-topological phase ($\delta_1 > \delta_2$). In (a) $\epsilon=0.1 \gamma$, (b) $\epsilon=0.2 \gamma$, (c) $\epsilon=0.3 \gamma$ and (d) $\epsilon=0.4 \gamma$. For all subplots $\delta_1=0.8 \gamma$, $\delta_2=0.5 \gamma$ and the length of the chain is $N=80$ unit cells.}
\label{T_vs_D_NonTop}

\end{figure}

In Fig. \ref{T_vs_D_Top} and Fig. \ref{T_vs_D_NonTop}, we depict the variation of $T$ with the strength of the NNN interaction ($\delta_3$) for the balanced loss-gain case.  Fig. \ref{T_vs_D_Top}  and  Fig. \ref{T_vs_D_NonTop} represent this variation for topological and non-topological cases, respectively. First, we emphasize the similarity between topological and non-topological phases with respect to $T$ oscillation. For the $\mathcal{PT}$ symmetric case ( $|\delta_1-\delta_2| > \epsilon$) of the isolated extended SSH Hamiltonian (\ref{ESSH}), $T$ shows oscillations with diminishing amplitudes with $\delta_3$ in both topological and non-topological phases as depicted in Fig. \ref{T_vs_D_Top} (a-b) and Fig. \ref{T_vs_D_NonTop} (a-b), respectively. However, in the $\mathcal{PT}$-broken phase ($|\delta_1-\delta_2| \leq \epsilon$), the amplitude of the oscillation first decreases, then increases slightly as $\delta_3$ increases further. Interestingly, we find that as $\delta_3$ increases, the maximum $T$ value frequently exceeds $1$ in all $\mathcal{PT}$-broken phases. This is expected because the energy levels with imaginary energy values, which appear as a ramification of $\mathcal{PT}$ breaking in the system, can cause $T$ to exceed $1$. One can interpret the $T>1$ case as a consequence of the addition of an effective current from the hypothetical reservoir whose degrees of freedom have been integrated out in the effective Hamiltonian of the extended SSH chain (\ref{ESSH}), and the net effect of the hypothetical reservoir is encoded within the complex onsite energy terms of the effective Hamiltonian \cite{TakaneJPSJ_2023}. Moreover, we notice in Fig. \ref{T_vs_D_Top} (a-d) and Fig. \ref{T_vs_D_NonTop} (a-d) that the $\delta_3$ range corresponding to $T=0$ at the initial part of the plots increases as $\epsilon$ increases.


Regarding the difference in transmission property in the topological and non-topological phases, we observe that the $T$ oscillation decreases very rapidly as $\delta_3$ increases in the case of the non-topological phase (Fig. \ref{T_vs_D_NonTop}), as compared to the topological phase (Fig. \ref{T_vs_D_Top}). We think that the effect of the non-Hermitian onsite energies on the density of states causes this difference in the transmission properties of the extended SSH chain in the topological phase compared to the same in the non-topological phase.

\begin{figure}[htb]
\centering{\includegraphics[width=1\linewidth]{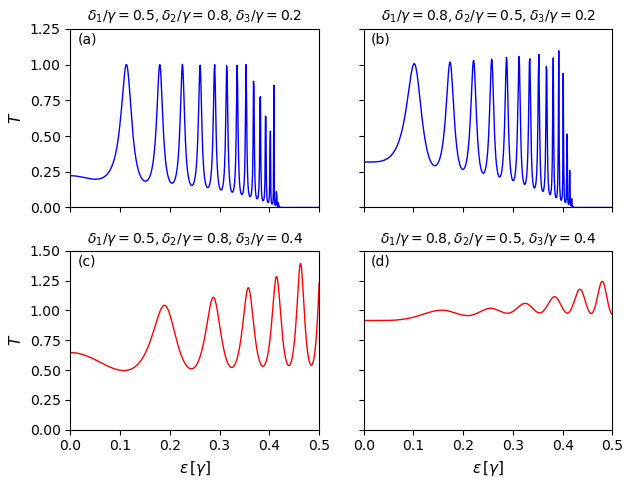}}
\caption{Variation of $T$ with $\epsilon$ for an incident plane wave with $E/\gamma=1.41421$ for (a) $\delta_1=0.5 \gamma, \delta_2=0.8 \gamma, \delta_3=0.2 \gamma$, (b) $\delta_1=0.8 \gamma, \delta_2=0.5 \gamma, \delta_3=0.2 \gamma$, (c) $\delta_1=0.5 \gamma, \delta_2=0.8 \gamma, \delta_3=0.4 \gamma$ and (d) $\delta_1=0.8 \gamma, \delta_2=0.5 \gamma, \delta_3=0.4 \gamma$. For all subplots the length of the chain is $N=100$ unit cells.}
\label{T_vs_ep}

\end{figure}

In Fig. \ref{T_vs_ep}, we study the variation of $T$ with non-Hermitian onsite energy ($\epsilon$) for varied strengths of NNN interaction but fixed energy incident wave. This study reveals an interesting interplay between the non-Hermiticity and the NNN interaction. We notice a clear distinction between the $T-\epsilon$ plots for to the smaller value of $\delta_3$ (Fig. \ref{T_vs_ep} (a-b)) and the larger value of $\delta_3$ (Fig. \ref{T_vs_ep} (c-d)), respectively. In Fig. \ref{T_vs_ep} (a-b), for $\delta_3=0.2 \gamma$, $T$ oscillates, but its amplitude decreases to zero as $\epsilon$ increases beyond some specific value which depends on the system's parameters. But in Fig. \ref{T_vs_ep} (c-d), for $\delta_3=0.4 \gamma$, amplitudes of the $T$ oscillations increase with $\epsilon$ and also go well above $1$ in the $\mathcal{PT}$ broken phase of the isolated Hamiltonian of the extended SSH chain (\ref{ESSH}) which is marked by the condition: $|\delta_1-\delta_2| < \epsilon $. For large $\delta_3$, such as $\delta_3/\gamma \approx 0.3, 0.4$ , we also observe qualitative difference in the $T$ oscillation pattern depending on whether the extended SSH chain is in the topological phase or the non-topological phase. In Fig. \ref{T_vs_ep} (c), the amplitude of the $T$ oscillation increases rapidly in the topological phase w.r.t $\epsilon$, but in Fig. \ref{T_vs_ep} (d), the amplitude of the $T$ oscillation increases slowly as $\epsilon$ increases while the chain is in the non-topological phase.

\subsubsection{Pure loss/gain terms  { \normalfont ($\epsilon_1=\epsilon_2=\epsilon$)}}

\begin{figure}[htb]
\centering{\includegraphics[width=1\linewidth]{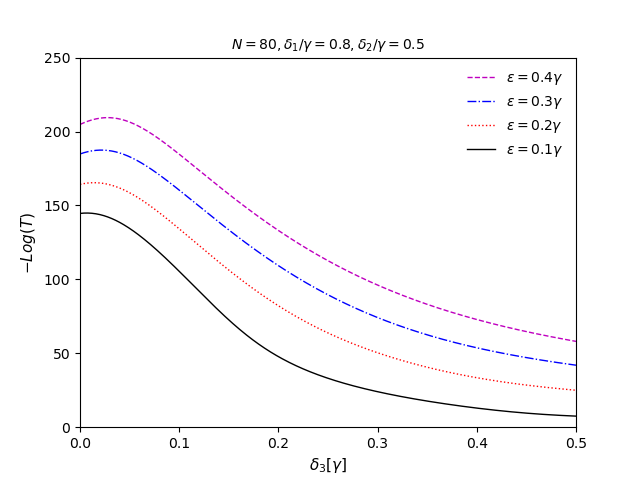}}
\caption{Variation of $-Log(T)$ with $\delta_3$ for an incident plane wave with $E/\gamma=1.41421$ while the extended SSH chain with Pure loss/gain term. Here $\delta_1=0.8 \gamma , \delta_2=0.5\gamma $ and the length of the chain is $N=80$ unit cells}
\label{T_vs_D_loss}
\end{figure}

This particular limit of non-Hermitian onsite terms actually mimics a situation in which the system is coupled to a hypothetical extended reservoir at each site, and this reservoir is responsible for injecting or leaking the probability current at each site. 
Clearly, this is a probability-nonconserving situation if we consider only the effective Hamiltonian. However, probability conservation is maintained once both the effective Hamiltonian and the hypothetical reservoir are considered together \cite{TakaneJPSJ_2023}. We plot the variation of $-Log(T)$ with $\delta_3$ for a pure loss/gain situation in Fig. \ref{T_vs_D_loss}. Here, we need to use logarithmic scale for $T$, because the value of $T$ is very small in such a pure loss/gain case. We find that the nature of the $-Log(T)$ vs $\delta_3$ curve remains similar for different values of onsite energies ($\epsilon$) however, this curve shifts upward as the strength of the onsite energy increases. Moreover, this behavior of the transmission coefficient remains almost the same in both the topological and non-topological phases.

\section{Summary and conclusion}
\label{sec:con}
In this work, we utilized the transmission coefficient to understand the transport characteristics of the one-dimensional extended SSH model under the influence of NNN interaction as well as non-Hermitian terms introduced via imaginary onsite energy. The Hermitian variant of the extended SSH chain with the NNN interaction shows strong modulation of the transmission coefficient with both the strength of the NNN interaction and the system size. For an extended SSH chain of fixed length, the magnitude of the transmission coefficient oscillates between a maximum value of $1$, and some relatively small non-zero values ($0.2 \sim 0.4$) as the strength of the NNN interaction varies.This oscillatory variation of $T$ is a unique property of the extended SSH chain, arising solely from long-range interactions. It may be viewed as the NNN interaction-induced transparency. In future, it will be interesting to investigate transport properties of a more generalized version of SSH, which might involve long-range interactions extended beyond the next-nearest-neighbor range. 

Another novel property we have observed is the strong dependence of transmission on the system size, i.e., the length of the chain. For some proper choices of the strength of the NNN interaction, intra and inter-molecular interaction, the transmission coefficient shows periodic oscillation between a transparent state ($T \approx 1$) and an opaque state ($T \approx 0$). This behavior can also be interpreted as system-size-dependent transparency. The reason for this transparency is destructive interference between the incident wave and the reflected waves, which are reflected from different intercell junctions. To the best of our knowledge, such transparency in an extended SSH chain has not been emphasized  before. 

In the presence of both the NNN interaction and balanced gain-loss non-Hermiticity, we observe interesting interplay between these two features.
In the $\mathcal{PT}$-symmetry case, the transmission coefficient exhibits weakly damped oscillatory behavior as the strength of the NNN interaction increases. On the other hand, for broken $\mathcal{PT}$-symmetry cases, the transmission coefficient undergoes oscillations whose amplitude initially decreases, then increases as the strength of the NNN interaction increases. The transmission coefficient can exceed $T=1$, once the energy eigenvalues become complex due to interactions with hypothetical reservoirs whose degrees of freedom have been traced out in the effective Hamiltonian, connected to each site.

We also observe that the transmission strongly depends on the strength of the  onsite potentials in the balanced loss-gain situation.  For weak NNN interaction, balanced gain-loss onsite energy causes damping in $T$ oscillations, which eventually reduces to zero as the onsite energy increases. But, for strong NNN interaction,  balanced gain-loss onsite energy boosts $T$ oscillations, with gradually increasing amplitude that frequently exceeds the $T = 1$ limit in the $\mathcal{PT}$-broken phase of the Hamiltonian (\ref{ESSH}). 

Therefore, the NNN interactions and the non-Hermitian terms in the Hamiltonian can be  selectively tuned to alter the transmission property of the extended SSH model. The main factor behind modulated transmission in this model is the NNN interaction, while the non-Hermitian terms also have a noticeable impact. This novel feature of an extended SSH chain, if realized in an experimental setup, could be useful in quantum information and quantum computing.

{\bf Acknowledgement:} NB sincerely thanks Yositake Takane for useful discussions on the numerical methods used in the simulation part of this work. NB acknowledges funding from DST-FIST programme.

\appendix
\setcounter{figure}{0}
\renewcommand\thefigure{A\arabic{figure}}
\section{Scattering Theory}
\label{App1}
Incoming plane waves and outgoing plane waves are connected by the matrix Green's function as follows,
\bea
\Psi = G X \, ,
\eea
where, outgoing waves is represented as 
\begin{widetext}
\beq
\Psi = \left( \psi_L(0), \psi_C(1), \psi_C(2), \dots, \psi_C(2N-1), \psi_C(2N), \psi_R(2N+1) \right)^T \, , 
\eeq
and, incoming wave, $X$ is represented by a $(2N+2)$ dimensional column matrix with elements, 
\beq
X_1=- 2i \gamma \sin(ka), X_2=\dots=X_{2N+2}=0 \,  .
\eeq
and, $G$ is $(2N+2) \times (2N+2)$ dimensional square matrix of following form:
\[
  G = 
  \left[ {\begin{array}{cccccccccccc}
    E-\gamma e^{ika} & -\Gamma & 0 & 0 & 0 & 0 & \cdots & 0 & 0 & 0 & 0\\
    -\Gamma & E-i\epsilon_1 & -\delta_1 & i\delta_3 & 0 & 0 & \cdots  & 0 & 0 & 0 & 0\\
     0 &  -\delta_1 & E-i\epsilon_2 & -\delta_2 & i\delta_3 & 0 &  \cdots   & 0 & 0 & 0 & 0\\
      0 &  -i\delta_3 & -\delta_2 & E-i\epsilon_1 &  -\delta_1 & i \delta_3 &  \cdots  & 0 & 0 & 0 &  0\\
    \vdots & \vdots & \ddots & \ddots & \ddots & \ddots &  \ddots & \vdots & \vdots &  \vdots &  \vdots\\
      0 & 0 & 0 & 0 & 0 & 0 &  \cdots & -i\delta_3  & -\delta_1 &  E-i\epsilon_2 & -\Gamma \\
    0 & 0 & 0 & 0 & 0 & 0  & \cdots & 0 & 0 & -\Gamma & E-\gamma e^{ika}\\
  \end{array} } \right]^{-1} \, ,
\]
where, $E = 2\gamma \cos(ka)$.
\end{widetext}


\bibliography{SSH_transport}

\end{document}